\documentclass[12pt]{article}
\usepackage{amsmath}
\usepackage{color}
\usepackage[unicode,bookmarks,bookmarksopen,bookmarksopenlevel=2,colorlinks,linkcolor=blue,citecolor=red]{hyperref}

\def\comment#1{}
\input{tcilatex}
\begin{document}

\title{Integrability of the Gibbons--Tsarev system}
\author{Maxim V. Pavlov \\
Department of Mathematical Physics,\\
P.N. Lebedev Physical Institute of Russian Academy of Sciences,\\
Moscow, Leninskij Prospekt, 53}
\date{}
\maketitle

\begin{abstract}
A new approach extracting multi-parametric hydrodynamic reductions for the
integrable hydrodynamic chains is presented. The Benney hydrodynamic chain
is considered.
\end{abstract}

\tableofcontents

\bigskip

\textit{keywords}: Poisson bracket, Hamiltonian structure, hydrodynamic
chain, Liouville coordinates\textit{.}

\bigskip


\section{Introduction}

The famous Benney hydrodynamic chain \cite{benney}%
\begin{equation}
A_{t}^{k}=A_{x}^{k+1}+kA^{k-1}A_{x}^{0}\text{, \ \ }k=0,1,...  \label{ch}
\end{equation}%
describes a fluid dynamics of a finite depth. This paper is dedicated to a
construction of particular solutions.

\textbf{1}. Let us suppose that each moment $A^{k}$ can be \textit{decomposed%
} on $N$ functions $f_{m(k)}(z)$ of a single variable ($N$ is arbitrary),
i.e.%
\begin{equation}
A^{k}=\overset{N}{\underset{m=1}{\dsum }}f_{m(k)}(a^{m}).  \label{decm}
\end{equation}

\textbf{2}. Let us substitute this ansatz (\ref{decm}) in (\ref{ch}) instead
all moments $A^{k}$ with \textit{variable} indices (except the first
equation in (\ref{ch})), and suppose that the \textit{summation} with
respect to the index $i$ can be removed from the hydrodynamic reduction%
\begin{equation*}
\overset{N}{\underset{i=1}{\dsum }}[f_{i(k)}^{\prime
}(a^{i})a_{t}^{i}-f_{i(k+1)}^{\prime
}(a^{i})a_{x}^{i}-kf_{i(k-1)}(a^{i})A_{x}^{0}]=0\text{,\ \ }k=1,2,...\text{,
\ \ }i=1,2,...,N.
\end{equation*}%
It means, that hydrodynamic type system%
\begin{equation*}
a_{t}^{i}=\frac{f_{i(k+1)}^{\prime }(a^{i})}{f_{i(k)}^{\prime }(a^{i})}%
a_{x}^{i}+\frac{kf_{i(k-1)}(a^{i})}{f_{i(k)}^{\prime }(a^{i})}A_{x}^{0}\text{%
, \ \ }k=1,2,...\text{, \ \ }i=1,2,...,N
\end{equation*}%
cannot explicitly depend on the discrete variable $k$. A solution of the
system%
\begin{equation*}
b_{i}(a^{i})=\frac{f_{i(k+1)}^{\prime }(a^{i})}{f_{i(k)}^{\prime }(a^{i})}%
\text{, \ \ \ }c_{i}(a^{i})=\frac{kf_{i(k-1)}(a^{i})}{f_{i(k)}^{\prime
}(a^{i})}
\end{equation*}%
is given by%
\begin{equation*}
f_{i(k)}(a^{i})=\epsilon _{i}\frac{b_{i}^{k+1}(a^{i})}{k+1}\text{, \ \ \ \ }%
\frac{1}{c_{i}(a^{i})}=b_{i}^{\prime }(a^{i}),
\end{equation*}%
where $\epsilon _{i}$ are arbitrary constants, and the functions $b_{i}(z)$
are not determined yet.

\textbf{3}. Let us suppose that $a^{i}$ are \textit{conservation law
densities}. Then $b_{i}(z)\equiv z$. The hydrodynamic reduction%
\begin{equation}
a_{t}^{i}=\left( \frac{(a^{i})^{2}}{2}+u\right) \text{, \ \ }i=1,2,...,N,
\label{zak}
\end{equation}%
where $u=A^{0}(\mathbf{a})$, is the so-called \textquotedblleft
symmetric\textquotedblright\ hydrodynamic type system (see \cite{algebra})
whose generating function of conservation laws%
\begin{equation}
p_{t}=\partial _{x}\left( \frac{p^{2}}{2}+u\right)  \label{on}
\end{equation}%
can be obtained by the formal replacement $a^{i}\rightarrow p$.

\textbf{Remark}: The above construction was derived from the whole Benney
hydrodynamic chain (\ref{ch}) excluding the first equation $%
A_{t}^{0}=A_{x}^{1}$. A substitution of the above formulae in this
conservation law implies the important constraint%
\begin{equation}
\sum \epsilon _{m}=0.  \label{strain}
\end{equation}

\textbf{4}. A consistency of (\ref{on}) with (\ref{zak}) implies the
so-called L\"{o}wner equation (see \cite{gt} and \cite{algebra})%
\begin{equation}
\partial _{i}p=\frac{\partial _{i}u}{a^{i}-p}\left( 1+\sum \frac{\partial
_{m}u}{a^{m}-p}\right) ^{-1},  \label{lev}
\end{equation}%
where $\partial _{i}\equiv \partial /\partial a^{i}$. The compatibility
condition $\partial _{i}(\partial _{k}p)=\partial _{k}(\partial _{i}p)$
leads to the Gibbons--Tsarev system (see \cite{gt} and \cite{algebra})%
\begin{equation}
(a^{i}-a^{k})\partial _{ik}u=\partial _{k}u\cdot \partial _{i}\delta
u-\partial _{i}u\cdot \partial _{k}\delta u\text{, \ }i\neq k,  \label{gt}
\end{equation}%
where $\delta \equiv \Sigma \partial /\partial a^{m}$. It is easy to see
that the above moment decomposition (\ref{decm})%
\begin{equation}
A^{0}(\mathbf{a})=\sum \epsilon _{m}a^{m}  \label{sym}
\end{equation}%
satisfies (\ref{gt}) \textit{without} restriction (\ref{strain}).

\textbf{Remark}: Suppose that symmetric hydrodynamic type system (\ref{gt})
possesses the conservation law $u_{t}=v_{x}$ only (see (\ref{ch})), where $%
v\equiv A^{1}(\mathbf{a})$. Then the compatibility condition $\partial
_{i}(\partial _{k}v)=\partial _{k}(\partial _{i}v)$ implies (\ref{gt})
again, where $\partial _{i}v=(a^{i}+\delta u)\partial _{i}u$. Thus, \textit{%
the existence of one extra conservation law leads to the existence of
infinitely many conservation laws} (see (\ref{on})).

\textbf{5}. L\"{o}wner equation (\ref{lev}) can be integrated once%
\begin{equation}
\lambda =p-\sum \epsilon _{m}\ln (p-a^{m})  \label{wat}
\end{equation}%
under substitution (\ref{sym}). $\lambda $ is an integration constant here.
Equation (\ref{wat}) of the Riemann surface (see \cite{Gib+Yu}, \cite%
{kod+water}) possesses an expansion ($\lambda \rightarrow \infty
,p\rightarrow \infty $)%
\begin{equation}
\lambda =p+\frac{A^{0}}{p}+\frac{A^{1}}{p^{2}}+\frac{A^{2}}{p^{3}}+...,
\label{ex}
\end{equation}%
where all higher moments are determined by the moment decomposition (see (%
\ref{decm}), (\ref{strain}), cf. (\ref{sym}))%
\begin{equation}
A^{k}=\frac{1}{k+1}\overset{N}{\underset{m=1}{\dsum }}\epsilon
_{m}(a^{m})^{k+1}\text{, \ }k=0,1,...  \label{ak}
\end{equation}%
Suppose that $\lambda $ is a function of $p$ and field variables $a^{k}$.
Then the Vlasov (collisionless Boltzmann) equation (see \cite{gibbons}, \cite%
{ops}, \cite{zakh})%
\begin{equation}
\lambda _{t}-p\lambda _{x}+\lambda _{p}u_{x}=0  \label{vlas}
\end{equation}%
is fulfilled due to (\ref{ch}) and (\ref{ex}). A substitution of the inverse
series%
\begin{equation}
p=\lambda -\frac{H_{0}}{\lambda }-\frac{H_{1}}{\lambda ^{2}}-\frac{H_{2}}{%
\lambda ^{3}}-...  \label{ser}
\end{equation}%
in (\ref{ex}) implies explicit expressions for all conservation law
densities $H_{k}(A^{0},A^{1},...,A^{k})$. For instance, $%
H_{0}=A^{0},H_{1}=A^{1},H_{2}=A^{2}+(A^{0})^{2},H_{3}=A^{3}+3A^{0}A^{1},...$
A substitution (\ref{ser}) in (\ref{on}) implies an infinite series of
conservation laws%
\begin{equation}
\partial _{t}H_{n}=\partial _{x}\left( H_{n+1}-\frac{1}{2}\overset{n-1}{%
\underset{m=0}{\dsum }}H_{m}H_{n-1-m}\right) ,\text{ \ }k=0,1,...
\label{cona}
\end{equation}

\textbf{6}. Hydrodynamic reductions (\ref{zak}) are semi-Hamiltonian (see 
\cite{tsar}) hydrodynamic type systems which can be written in the diagonal
form (see \cite{gt}, \cite{ops})%
\begin{equation}
r_{t}^{i}=p^{i}(\mathbf{r})r_{x}^{i}\text{, \ \ \ }i=1,2,...,N  \label{rim}
\end{equation}%
by virtue of invertible point transformations $a^{k}(\mathbf{r})$.
Consistency of (\ref{rim}) with (\ref{on}) leads to the L\"{o}wner equation (%
$\partial _{i}\equiv \partial /\partial r^{i}$)%
\begin{equation*}
\partial _{i}p=\frac{\partial _{i}u}{p^{i}-p},
\end{equation*}%
written via Riemann invariants $r^{k}$ (cf. (\ref{lev})) whose compatibility
condition $\partial _{i}(\partial _{k}p)=\partial _{k}(\partial _{i}p)$
implies the Gibbons--Tsarev system (cf. (\ref{gt}))%
\begin{equation}
\partial _{i}p^{k}=\frac{\partial _{i}u}{p^{i}-p^{k}}\text{, \ \ \ }\partial
_{ik}u=2\frac{\partial _{i}u\cdot \partial _{k}u}{(p^{i}-p^{k})^{2}}\text{,
\ \ \ }i\neq k,  \label{got}
\end{equation}%
originally derived in \cite{gt}.

\textbf{7}. An arbitrary $N$ component hydrodynamic reduction (\ref{rim})
can be written in the \textit{combined} form (see (\ref{zak}) and (\ref{cona}%
))%
\begin{eqnarray}
a_{t}^{k} &=&\partial _{x}\left( \frac{(a^{k})^{2}}{2}+H_{0}\right) \text{,
\ \ }k=1,2,...,K,  \notag \\
&&  \notag \\
\partial _{t}H_{n} &=&\partial _{x}\left( H_{n+1}-\frac{1}{2}\overset{n-1}{%
\underset{m=0}{\dsum }}H_{m}H_{n-1-m}\right) ,\text{ \ }n=0,1,...,M-2,
\label{tri} \\
&&  \notag \\
\partial _{t}H_{M-1} &=&\partial _{x}\left( H_{M}(\mathbf{a},\mathbf{H})-%
\frac{1}{2}\overset{M-2}{\underset{m=0}{\dsum }}H_{m}H_{M-2-m}\right) , 
\notag
\end{eqnarray}%
where $N=K+M$, and $H_{M}(\mathbf{a},\mathbf{H})$ is some function
determined from a consistency of the above hydrodynamic type system and the
generating function of conservation laws (\ref{on}), where $u\equiv H_{0}$.

\textbf{Remark}: A simplest choice $H_{M}(\mathbf{a},\mathbf{H})=\Sigma
\epsilon _{m}a^{m}$ extracts hydrodynamic reductions (\ref{tri}) equipped by
local Hamiltonian structures of the Dubrovin--Novikov type (see the next
Section).

Gibbons--Tsarev system (\ref{got}) describes integrable hydrodynamic
reductions (semi-Hamiltonian hydrodynamic type systems) of integrable Benney
hydrodynamic chain (\ref{ch}). Thus, we believe that the Gibbons--Tsarev
system is also integrable. At least, we know, that the Gibbons--Tsarev
system possesses infinitely many solutions parameterized by $N$ arbitrary
functions of a single variable (see \cite{gt}). All known solutions were
found in \cite{bogdan}, \cite{gibbons}, \cite{krich}, \cite{kuper}, \cite%
{zakh}. Nevertheless, a construction of such solutions is an open problem
now. By this reason, we try to suggest the new approach allowing to seek
most general families of particular solutions. This paper is devoted to a
description of hydrodynamic reductions (\ref{tri}), where $H_{M}(\mathbf{a},%
\mathbf{H})$ depends on all lower field variables $H_{n}$ ($n=0,1,...,M-1)$
and a sole function $\Delta =\Sigma f_{m}(a^{m})$.

This paper is organized in the following way. In Section 2, hydrodynamic
reductions equipped by local Hamiltonian structures of the Dubrovin--Novikov
type are determined by the special choice $H_{M}=\Sigma \epsilon _{m}a^{m}$.
In Section 3, $N+2$ parametric family of solutions $u(\mathbf{a})$ of the
Gibbons--Tsarev system is found for the first special sub-case $%
H_{0}=f_{0}(\Delta )$. A corresponding equation of the Riemann surface and
corresponding two-parametric family of hydrodynamic chains are constructed.
In Section 4, the second special sub-case $H_{1}=f_{1}(\Delta )$ is
considered. Corresponding hydrodynamic chain and the Riemann surface are
presented. In Conclusion, a generalization of the approach presented in this
paper is discussed.

\section{Hamiltonian reductions}

The Benney hydrodynamic chain belongs to the class of Egorov hydrodynamic
chains (see details in \cite{egor}, \cite{Maks+Tsar}), which possess a pair
of conservation laws $a_{t}=b_{x},b_{t}=c_{x}$. In the case of the Benney
hydrodynamic chain, $a=H_{0},b=H_{1},c=H_{2}-H_{0}^{2}/2$ (see (\ref{cona}%
)). All commuting flows of any integrable Egorov hydrodynamic chain are
integrable Egorov hydrodynamic chains too. A generating function of such
Egorov's pair is given by the so-called dispersionless Hirota equations
(see, for instance, \cite{Yuji+Carroll})%
\begin{equation}
\partial _{\tau (\zeta )}H_{0}=\partial _{x}p(\zeta )\text{, \ \ \ \ }%
\partial _{\tau (\zeta )}p(\lambda )=\partial _{x}\ln [p(\lambda )-p(\zeta
)],  \label{geg}
\end{equation}%
where $p(\lambda )$ is a generating function of conservation law densities
(see (\ref{on}) and (\ref{ser})) and $\partial _{\tau (\zeta )}$ is a formal
operator (see, for instance, in \cite{egor}). Thus, the generating function
of commuting hydrodynamic reductions (they are obtained by a formal
replacement $p(\lambda )\rightarrow a^{i}$ exactly as in (\ref{on}))%
\begin{equation}
a_{\tau (\zeta )}^{i}=\partial _{x}\ln (p(\zeta )-a^{i})  \label{red}
\end{equation}%
possesses an infinite series of conservation laws (they can be obtained by a
substitution of (\ref{ser}) in (\ref{geg}))%
\begin{equation*}
\partial _{\tau (\zeta )}H_{0}=\partial _{x}p(\zeta )\text{, \ \ }\partial
_{\tau (\zeta )}H_{1}=\partial _{x}\left( \frac{p^{2}(\zeta )}{2}%
+H_{0}\right) \text{,\ \ \ }\partial _{\tau (\zeta )}H_{2}=\partial
_{x}\left( \frac{p^{3}(\zeta )}{3}+H_{0}p(\zeta )+H_{1}\right) ,...
\end{equation*}

\textbf{Remark}: A substitution of (\ref{ser}) for $p(\zeta )$ and the
formal series%
\begin{equation}
\partial _{\tau (\zeta )}=-\frac{1}{\zeta }\partial _{t^{0}}-\frac{1}{\zeta
^{2}}\partial _{t^{1}}-\frac{1}{\zeta ^{3}}\partial _{t^{2}}-...  \label{ga}
\end{equation}%
in (\ref{geg}) leads to an infinite series of generating functions (cf. (\ref%
{on}))%
\begin{equation*}
p_{t^{0}}=p_{x}\text{, \ \ \ \ }p_{t^{1}}=\partial _{x}\left( \frac{p^{2}}{2}%
+H_{0}\right) \text{,\ \ \ }p_{t^{2}}=\partial _{x}\left( \frac{p^{3}}{3}%
+H_{0}p+H_{1}\right) ,...
\end{equation*}%
A substitution of (\ref{ser}) for $p(\zeta )$ and (\ref{ga}) in (\ref{red})
leads to the infinite series of commuting hydrodynamic reductions%
\begin{equation*}
a_{t^{1}}^{k}=\partial _{x}\left( \frac{(a^{k})^{2}}{2}+H_{0}(\mathbf{a}%
)\right) \text{,\ \ \ }a_{t^{2}}^{k}=\partial _{x}\left( \frac{(a^{k})^{3}}{3%
}+a^{k}H_{0}(\mathbf{a})+H_{1}(\mathbf{a})\right) ,...
\end{equation*}%
Thus, the generalized hodograph method (established by S.P. Tsarev in \cite%
{tsar} and adopted for an arbitrary set of commuting hydrodynamic type
systems by J. Gibbons and Yu. Kodama in \cite{Gib+Yu}, \cite{kod+water})
implies an infinite series of particular solutions determined by the
generating function%
\begin{equation*}
xda^{k}+td\left( \frac{(a^{k})^{2}}{2}+H_{0}(\mathbf{a})\right) +yd\left( 
\frac{(a^{k})^{3}}{3}+a^{k}H_{0}(\mathbf{a})+H_{1}(\mathbf{a})\right) =d\ln
(p-a^{k})\text{, \ \ }k=1,2,...,N
\end{equation*}%
for the Benney hydrodynamic chain as well as for the Khohlov--Zabolotzkaya
system (see, for instance, \cite{krich})%
\begin{equation*}
u_{t}=v_{x}\text{, \ \ \ \ \ }u_{y}=v_{t}+uu_{x},
\end{equation*}%
where $x=t^{0},t=t^{1},y=t^{2}$.

Let us consider a set of hydrodynamic reductions (\ref{tri}) extracted by
the constraint $H_{M}=\Sigma \epsilon _{m}a^{m}$.

\textbf{1}. In the case of the waterbag reduction (\ref{sym}), $H_{0}=\Sigma
\epsilon _{m}a^{m}$. Thus, the simple computation (see (\ref{red}))%
\begin{equation*}
\partial _{\tau (\zeta )}H_{0}\equiv \partial _{\tau (\zeta )}\left( \sum
\epsilon _{m}a^{m}\right) =\partial _{x}\left( \sum \epsilon _{m}\ln
(p-a^{m})\right) =\partial _{x}p(\zeta )
\end{equation*}%
leads to equation of the Riemann surface (\ref{wat}). The waterbag
hydrodynamic reduction (see (\ref{zak}) and (\ref{wat})) possesses the local
Hamiltonian structure of the Dubrovin--Novikov type%
\begin{equation*}
a_{t}^{i}=\frac{1}{\epsilon _{i}}\partial _{x}\frac{\partial H_{2}}{\partial
a^{i}},
\end{equation*}%
where the momentum and the Hamiltonian densities are given, respectively, by%
\begin{equation*}
H_{1}=\frac{1}{2}\sum \epsilon _{m}(a^{m})^{2}\text{, \ \ }H_{2}=\frac{1}{6}%
\sum \epsilon _{m}(a^{m})^{3}+\frac{1}{2}\left( \sum \epsilon
_{m}a^{m}\right) ^{2}.
\end{equation*}

\textbf{2}. Suppose, in comparison with the waterbag case, that $%
H_{1}=\Sigma \epsilon _{m}a^{m}$. Then (see (\ref{red}))%
\begin{equation*}
\partial _{\tau (\zeta )}H_{1}\equiv \partial _{\tau (\zeta )}\left( \sum
\epsilon _{m}a^{m}\right) =\partial _{x}\left( \sum \epsilon _{m}\ln
(p(\zeta )-a^{m})\right) =\partial _{x}\left( \frac{p^{2}(\zeta )}{2}%
+H_{0}\right) 
\end{equation*}%
leads to the equation of the Riemann surface%
\begin{equation*}
\lambda =\frac{p^{2}}{2}+H_{0}-\sum \epsilon _{m}\ln (p-a^{m})
\end{equation*}%
associated with $K+1$ component hydrodynamic type system%
\begin{equation*}
a_{t}^{k}=\partial _{x}\left( \frac{(a^{k})^{2}}{2}+H_{0}\right) \text{, \ \ 
}\partial _{t}H_{0}=\partial _{x}\left( \sum \epsilon _{m}a^{m}\right) ,
\end{equation*}%
which has the local Hamiltonian structure%
\begin{equation*}
a_{t}^{i}=\frac{1}{\epsilon _{i}}\partial _{x}\frac{\partial H_{3}}{\partial
a^{i}}\text{, \ \ }\partial _{t}H_{0}=\partial _{x}\frac{\partial H_{3}}{%
\partial H_{0}},
\end{equation*}%
where the momentum and the Hamiltonian densities are given, respectively, by%
\begin{equation*}
H_{2}=\frac{1}{2}\sum \epsilon _{m}(a^{m})^{2}+\frac{1}{2}H_{0}^{2}\text{, \
\ }H_{3}=\frac{1}{6}\sum \epsilon _{m}(a^{m})^{3}+H_{0}\sum \epsilon
_{m}a^{m}.
\end{equation*}

\textbf{3}. Suppose now, that $H_{2}=\Sigma \epsilon _{m}a^{m}$. Then (see (%
\ref{red}))%
\begin{equation*}
\partial _{\tau (\zeta )}H_{2}\equiv \partial _{\tau (\zeta )}\left( \sum
\epsilon _{m}a^{m}\right) =\partial _{x}\left( \sum \epsilon _{m}\ln
(p(\zeta )-a^{m})\right) =\partial _{x}\left( \frac{p^{3}(\zeta )}{3}%
+H_{0}p(\zeta )+H_{1}\right) 
\end{equation*}%
leads to the equation of the Riemann surface%
\begin{equation*}
\lambda =\frac{p^{3}}{2}+H_{0}p+H_{1}-\sum \epsilon _{m}\ln (p-a^{m}),
\end{equation*}%
associated with $K+2$ component hydrodynamic type system

\begin{equation*}
a_{t}^{k}=\partial _{x}\left( \frac{(a^{k})^{2}}{2}+H_{0}\right) \text{, \ \ 
}\partial _{t}H_{0}=\partial _{x}H_{1}\text{, \ \ \ }\partial
_{t}H_{1}=\partial _{x}\left( \sum \epsilon _{m}a^{m}-\frac{H_{0}^{2}}{2}%
\right) ,
\end{equation*}%
which has the local Hamiltonian structure%
\begin{equation*}
a_{t}^{i}=\frac{1}{\epsilon _{i}}\partial _{x}\frac{\partial H_{4}}{\partial
a^{i}}\text{, \ \ }\partial _{t}H_{0}=\partial _{x}\frac{\partial H_{4}}{%
\partial H_{1}}\text{, \ \ }\partial _{t}H_{1}=\partial _{x}\frac{\partial
H_{4}}{\partial H_{0}},
\end{equation*}%
where the momentum and the Hamiltonian densities are given, respectively, by%
\begin{equation*}
H_{3}=\frac{1}{2}\sum \epsilon _{m}(a^{m})^{2}+H_{0}H_{1}\text{, \ \ }H_{4}=%
\frac{1}{6}\sum \epsilon _{m}(a^{m})^{3}+\frac{1}{2}H_{1}^{2}-\frac{1}{6}%
H_{0}^{3}+H_{0}\sum \epsilon _{m}a^{m}.
\end{equation*}

All other hydrodynamic reductions of the Benney hydrodynamic chain
possessing local Hamiltonian structures%
\begin{equation}
a_{t}^{i}=\frac{1}{\epsilon _{i}}\partial _{x}\frac{\partial H_{M+2}}{%
\partial a^{i}}\text{, \ \ }\partial _{t}H_{k}=\partial _{x}\frac{\partial
H_{M+2}}{\partial H_{M-k}}  \label{ham}
\end{equation}%
are associated with equations of the Riemann surfaces%
\begin{equation}
\lambda =\frac{p^{M+1}}{M+1}+H_{0}p^{M-1}+H_{1}p^{M-2}+P_{M-3}(\mathbf{H}%
,p)-\sum \epsilon _{m}\ln (p-a^{m}),  \label{i}
\end{equation}%
where $P_{M-3}(\mathbf{H},p)$ is a polynomial of a degree $M-3$ with respect
to $p$, whose coefficients depend on $H_{0},...,H_{M-1}$. In such a case,
the moment decomposition (cf. (\ref{ak}))%
\begin{equation}
\tilde{A}^{k}=\frac{1}{k+1-M}\overset{N}{\underset{m=1}{\dsum }}\epsilon
_{m}(a^{m})^{k+1-M}\text{, \ }k=M,M+1,...  \label{ok}
\end{equation}%
reduces the Benney hydrodynamic chain%
\begin{equation}
\tilde{A}_{t}^{0}=\tilde{A}_{x}^{1},\text{ \ \ \ \ }\tilde{A}_{t}^{k}=\tilde{%
A}_{x}^{k+1}+(k-M)\tilde{A}^{k-1}\tilde{A}_{x}^{0}\text{, \ \ }k=1,2,...
\label{mod}
\end{equation}%
to (\ref{ham}). These hydrodynamic reductions were found by the so-called
\textquotedblleft symmetry constraint\textquotedblright\ approach (see \cite%
{bogdan}). The moments $\tilde{A}^{k}$ are connected with $A^{k}$ by
invertible triangular point transformations $\tilde{A}%
^{k}(A^{0},A^{1},...A^{k})$. It is easy to see comparing equations of the
Riemann surfaces (cf. (\ref{ex}) and (\ref{i}))%
\begin{equation*}
\lambda =\frac{1}{M+1}\left( p+\overset{\infty }{\underset{k=0}{\dsum }}%
\frac{A^{k}}{p^{k+1}}\right) ^{M+1}=\frac{p^{M+1}}{M+1}+p^{M}\overset{\infty 
}{\underset{k=0}{\dsum }}\frac{\tilde{A}^{k}}{p^{k+1}}.
\end{equation*}

\textbf{Remark}: More complicated hydrodynamic reductions (cf. (\ref{tri}))%
\begin{eqnarray}
a_{t}^{k} &=&\partial _{x}\left( \frac{(a^{k})^{2}}{2}+H_{0}\right) \text{,
\ \ }k=1,2,...,K,  \notag \\
&&  \notag \\
\partial _{t}H_{n} &=&\partial _{x}\left( H_{n+1}-\frac{1}{2}\overset{n-1}{%
\underset{m=0}{\dsum }}H_{m}H_{n-1-m}\right) ,\text{ \ }n=0,1,...,M-2,
\label{j} \\
&&  \notag \\
\partial _{t}H_{M-1} &=&\partial _{x}\left( H_{M}(\Delta ,\mathbf{H})-\frac{1%
}{2}\overset{M-2}{\underset{m=0}{\dsum }}H_{m}H_{M-2-m}\right) ,  \notag
\end{eqnarray}%
where $\Delta =\Sigma f_{k}(a^{k})$, can be associated with different
Riemann mappings $\lambda (p)$. Vlasov equation (\ref{vlas}) is invariant
under an arbitrary scaling $\tilde{\lambda}(\lambda )$. Let us consider in
such a case, the substitution (instead of (\ref{ex}))%
\begin{equation}
\lambda =f\left( p+\frac{A^{0}}{p}+\frac{A^{1}}{p^{2}}+\frac{A^{2}}{p^{3}}%
+...\right) ,  \label{fan}
\end{equation}%
where $f(p)$ is an arbitrary entire function. A corresponding Taylor
expansion%
\begin{equation*}
\lambda =f(p)+\frac{1}{p}\left( A^{0}+\frac{A^{1}}{p}+\frac{A^{2}}{p^{2}}%
+...\right) f^{\prime }(p)+\frac{1}{2p^{2}}\left( A^{0}+\frac{A^{1}}{p}+%
\frac{A^{2}}{p^{2}}+...\right) ^{2}f^{\prime \prime }(p)+...
\end{equation*}%
can be reduced to the form%
\begin{equation*}
\lambda =f(p)+\tilde{f}(p)\cdot \left( \tilde{A}^{0}+\frac{\tilde{A}^{1}}{p}+%
\frac{\tilde{A}^{2}}{p^{2}}+...\right) ,
\end{equation*}%
where $\tilde{A}^{k}(A^{0},A^{1},...,A^{k})$ are polynomials, and $\tilde{f}%
(p)$ is some function. Indeed, a substitution of this ansatz in the Vlasov
equation leads to the Benney hydrodynamic chain written in the form%
\begin{equation*}
\tilde{A}_{t}^{k}=\tilde{A}_{x}^{k+1}-[\beta _{-1}\tilde{A}^{k+1}+\beta _{0}%
\tilde{A}^{k}+(\beta _{1}+1-k)\tilde{A}^{k-1}+\beta _{2}\tilde{A}%
^{k-2}+...+\beta _{k}\tilde{A}^{0}+\alpha _{k}]\frac{\tilde{A}_{x}^{0}}{%
\alpha _{-1}+\beta _{-1}\tilde{A}^{0}},
\end{equation*}%
where%
\begin{equation*}
\frac{f^{\prime }(p)}{\tilde{f}(p)}=\alpha _{-1}p+\alpha _{0}+\frac{\alpha
_{1}}{p}+\frac{\alpha _{2}}{p^{2}}+...\text{, \ \ \ }\frac{\tilde{f}^{\prime
}(p)}{\tilde{f}(p)}=\beta _{-1}p+\beta _{0}+\frac{\beta _{1}}{p}+\frac{\beta
_{2}}{p^{2}}+...
\end{equation*}%
If, for instance, $f^{\prime }(p)=p^{M}$, then the Benney hydrodynamic chain
($\alpha _{-1}=1$, $\beta _{1}=M-1$; all other coefficients $\alpha _{k}$
and $\beta _{k}$ vanish) can be written in form (\ref{mod}). We believe that 
\textit{all moments} $\tilde{A}^{k}$ \textit{(except first }$M$\textit{\
moments }$\tilde{A}^{0},\tilde{A}^{0},...,\tilde{A}^{M-1}$\textit{) in the
Benney hydrodynamic chain written in the above form can be expressed via
field variables} $a^{i}$ \textit{as (cf. }(\ref{decm})\textit{)}%
\begin{equation}
\tilde{A}^{k}=\overset{N}{\underset{m=1}{\dsum }}f_{m(k)}(a^{m})\text{, \ \ }%
k=M,M+1,...,  \label{reg}
\end{equation}%
\textit{for some set of constants }$\alpha _{k}$ \textit{and} $\beta _{k}$%
\textit{, where the functions} $f_{m(k)}(z)$ \textit{are uniquely determined
by the sole function} $f(z)$ (\textit{see} (\ref{fan})). Indeed, a
substitution of this ansatz in (\ref{mod}) immediately implies (\ref{ok}).
Several such sub-cases are considered in next two Sections.

\section{Simplest solutions of the Gibbons--Tsarev system}

The first non-trivial case in this paper is given by $N$ component
hydrodynamic reduction (\ref{zak}), where $u$ is a solution of
Gibbons--Tsarev system (\ref{gt}). Let us suppose that $u$ depends on a sole
function $\Delta =\Sigma f_{k}(a^{k})$ only. In such a case, three
distinguished sub-cases can be extracted.

\textbf{I}. Gibbons--Tsarev system (\ref{gt}) admits $N$ parametric solution
(\ref{sym}). This sub-case is well known, and already is considered with
constraint (\ref{strain}) in previous Sections. Let us briefly consider this
sub-case without constraint (\ref{strain}). Since a solution of the linear
PDE (\ref{vlas}) is determined up to an arbitrary function $\tilde{\lambda}%
(\lambda )$, let us choose a new function $\lambda $ such that (see (\ref%
{wat}))%
\begin{equation*}
\lambda -\epsilon \ln \lambda =p-\sum \epsilon _{m}\ln (p-a^{m}),
\end{equation*}%
where $\epsilon \equiv \Sigma \epsilon _{m}$. A substitution of (\ref{ser})
instead $\lambda $ in the above formula leads to explicit expressions of
conservation law densities $H_{n}$ via moments $A^{k}$ (see (\ref{ak})).
Then the Benney hydrodynamic chain%
\begin{equation*}
A_{t}^{0}=\partial _{x}(A^{1}+\epsilon A^{0})\text{, \ \ \ }%
A_{t}^{k}=A_{x}^{k+1}+kA^{k-1}A_{x}^{0}\text{, \ }k=1,2,...
\end{equation*}%
is equipped by the \textit{deformed} Kupershmidt--Manin Poisson bracket (cf. 
\cite{kuper})%
\begin{equation*}
\{A^{0},A^{0}\}=\epsilon \delta ^{\prime }(x-x^{\prime })\text{, \ \ }%
\{A^{k},A^{n}\}=[kA^{k+n-1}\partial _{x}+n\partial _{x}A^{k+n-1}]\delta
(x-x^{\prime })\text{, \ }k+n>0.
\end{equation*}

\textbf{II}. Let us look for a solution of Gibbons--Tsarev system (\ref{gt})
in the more general form $u=\Sigma f_{m}(a^{m})$, where $f_{k}(a^{k})$ are
not yet determined functions. A substitution of this ansatz in (\ref{gt})
leads to the new $N+1$ parametric solution%
\begin{equation*}
u=\frac{1}{\epsilon _{0}}\sum \epsilon _{m}e^{\epsilon _{0}a^{m}},
\end{equation*}%
where $\epsilon _{0}$ and $\epsilon _{k}$ are arbitrary constants. A
corresponding equation of the Riemann surface is given by%
\begin{equation}
\lambda =-\frac{1}{\epsilon _{0}}e^{-\epsilon _{0}p}-\sum \epsilon _{m}\int
e^{-\epsilon _{0}(p-a^{m})}d\ln (p-a^{m}).  \label{exp}
\end{equation}%
Hydrodynamic type system (\ref{zak}) can be rewritten as the Benney
hydrodynamic chain written in the form%
\begin{equation*}
A_{t}^{k}=A_{x}^{k+1}+(\epsilon _{0}A^{k}+kA^{k-1})A_{x}^{0},
\end{equation*}%
where%
\begin{equation*}
dA^{k}=\sum \epsilon _{m}(a^{m})^{k}e^{\epsilon _{0}a^{m}}da^{m}.
\end{equation*}%
Then (\ref{exp}) (instead (\ref{ex})) reduces to the form ($p\rightarrow
\infty $)%
\begin{equation*}
\lambda =e^{-\epsilon _{0}p}\left( -\frac{1}{\epsilon _{0}}+\overset{\infty }%
{\underset{m=0}{\dsum }}\frac{A^{n}}{p^{n+1}}\right) .
\end{equation*}

\textbf{III}. Let us look for a solution of Gibbons--Tsarev system (\ref{gt}%
) in the form $u(\Delta )$, where $\Delta =\Sigma f_{m}(a^{m})$, where $%
f_{k}(a^{k})$ are not yet determined functions. A substitution of this
ansatz \ in (\ref{gt}) leads to the new $N+2$ parametric solution%
\begin{equation*}
u=\frac{1}{\sigma }\ln (\sigma \Delta +\xi )\text{, \ \ \ \ \ }f_{k}^{\prime
}(a^{k})=\epsilon _{k}\exp \left( \epsilon _{0}a^{k}-\frac{\sigma (a^{k})^{2}%
}{2}\right) 
\end{equation*}%
where $\sigma ,\xi ,\epsilon _{0}$ and $\epsilon _{k}$ are arbitrary
constants (if $\sigma \neq 0$, the constant $\xi $ is removable by an
appropriate shift $A^{0}\rightarrow A^{0}-\xi /\sigma $, i.e. just $N+2$
independent arbitrary constants). A corresponding equation of the Riemann
surface can be found in quadratures%
\begin{equation*}
d\lambda =(\sigma \Delta +\xi )e^{\frac{\sigma }{2}p^{2}-\epsilon
_{0}p}dp-e^{\frac{\sigma }{2}p^{2}-\epsilon _{0}p}\sum \epsilon
_{m}e^{\epsilon _{0}a^{m}-\frac{\sigma }{2}(a^{m})^{2}}d\ln (p-a^{m}).
\end{equation*}

Let us introduce an infinite set of moments $A^{k}(\mathbf{a})$ determined
by quadratures%
\begin{equation*}
dA^{k}=\sum \epsilon _{m}(a^{m})^{k}e^{\epsilon _{0}a^{m}-\frac{\sigma }{2}%
(a^{m})^{2}}da^{m}.
\end{equation*}%
In such a case, hydrodynamic type system can be written as the Benney
hydrodynamic chain (cf. (\ref{ch}) and (\ref{cona})) written in the form%
\begin{equation*}
A_{t}^{k}=A_{x}^{k+1}+(-\sigma A^{k+1}+\epsilon _{0}A^{k}+kA^{k-1})\frac{%
A_{x}^{0}}{\sigma A^{0}+\xi }.
\end{equation*}%
Then above equation of the Riemann surface reduces to the form ($%
p\rightarrow \infty $)%
\begin{equation*}
\lambda =\xi \int e^{\frac{\sigma }{2}p^{2}-\epsilon _{0}p}dp+e^{\frac{%
\sigma }{2}p^{2}-\epsilon _{0}p}\overset{\infty }{\underset{n=0}{\sum }}%
\frac{A^{n}}{p^{n+1}}.
\end{equation*}

\section{Second and third particular cases}

In this Section, two extra multi-parametric generalizations (see (\ref{j}))
of hydrodynamic reductions equipped by local Hamiltonian structures of the
Dubrovin--Novikov type (see Section 2) are described.

\textbf{1}. $K+1$ component hydrodynamic reduction (cf. (\ref{zak}) and (\ref%
{tri}))%
\begin{equation}
u_{t}=v_{x}\text{, \ \ }a_{t}^{k}=\left( \frac{(a^{k})^{2}}{2}+u\right) 
\text{, \ \ }k=1,2,...,K,  \label{ma}
\end{equation}%
where $v\equiv H_{1}(\mathbf{a},u)$, is a semi-Hamiltonian hydrodynamic type
system if at least one extra conservation law (see (\ref{cona}), $n=1$)%
\begin{equation*}
v_{t}=\partial _{x}\left( w-\frac{u^{2}}{2}\right) ,
\end{equation*}%
where $w\equiv H_{2}(\mathbf{a},u)$, exists. In such a case, the
compatibility conditions $\partial _{i}(\partial _{k}w)=\partial
_{k}(\partial _{i}w)$ and $\partial _{k}(\partial _{u}w)=\partial
_{u}(\partial _{k}w)$ lead to the Gibbons--Tsarev system (cf. (\ref{gt}))%
\begin{eqnarray*}
(a^{i}-a^{k})\partial _{ik}v &=&\partial _{k}v\cdot \partial _{iu}v-\partial
_{i}v\cdot \partial _{ku}v\text{, \ }i\neq k, \\
&& \\
\partial _{k}v\cdot \partial _{uu}v &=&(\partial _{u}v-a^{k})\partial
_{ku}v+\partial _{k}\delta v,
\end{eqnarray*}%
where $\partial _{u}w=\delta v+(\partial _{u}v)^{2}+u$ and $\partial
_{i}w=(\partial _{u}v+a^{i})\partial _{i}v$. A consistency of (\ref{on}) and
(\ref{ma}) leads to the L\"{o}wner equations (cf. (\ref{lev}))%
\begin{equation*}
\partial _{i}p=\frac{\partial _{i}v}{a^{i}-p}\left( p-\partial _{u}v+\sum 
\frac{\partial _{m}v}{a^{m}-p}\right) ^{-1}\text{, \ \ \ }\partial
_{u}p=-\left( p-\partial _{u}v+\sum \frac{\partial _{m}v}{a^{m}-p}\right)
^{-1}.
\end{equation*}

Suppose (like in the previous Section) that $v$ depends on $u$ and a sole
function $\Delta =\Sigma f_{k}(a^{k})$ only. This ansatz reduces the above
Gibbons--Tsarev system to%
\begin{equation*}
\partial _{\Delta \Delta }v=0\text{, \ \ }\partial _{uu}v=(\partial
_{u}v-a^{k})\frac{\partial _{u\Delta }v}{\partial _{\Delta }v}+\frac{%
f_{k}^{\prime \prime }(a^{k})}{f_{k}^{\prime }(a^{k})},
\end{equation*}%
whose solution is given by%
\begin{equation*}
v=(\beta \Delta +\alpha )e^{-\epsilon _{0}u}+\frac{\gamma }{\epsilon _{0}}u%
\text{, \ \ \ }\ln f_{k}^{\prime }=\ln \epsilon _{k}+\gamma a^{k}-\frac{%
\epsilon _{0}}{2}(a^{k})^{2},
\end{equation*}%
where $\alpha ,\beta ,\gamma ,\epsilon _{0}$ and $\epsilon _{k}$ are
constants.

Then these L\"{o}wner equations can be integrated once, and equation of the
Riemann surface can be found in quadratures%
\begin{equation*}
d\lambda =\frac{1}{\epsilon _{0}}de^{\frac{\epsilon _{0}}{2}p^{2}-\gamma
p+\epsilon _{0}u}+\epsilon _{0}(\beta \Delta +\alpha )e^{\frac{\epsilon _{0}%
}{2}p^{2}-\gamma p}dp-\beta e^{\frac{\epsilon _{0}}{2}p^{2}-\gamma p}\sum
f_{m}^{\prime }(a^{m})d\ln (p-a^{m}).
\end{equation*}%
Introducing moments $A^{k}$ by their derivatives%
\begin{equation*}
dA^{k}=\sum (a^{m})^{k-1}f_{m}^{\prime }(a^{m})da^{m}\text{, \ }k=1,2,...,
\end{equation*}%
the above equation reduces to%
\begin{equation*}
\lambda =\frac{1}{\epsilon _{0}}e^{\frac{\epsilon _{0}}{2}p^{2}-\gamma
p+\epsilon _{0}u}+\alpha \epsilon _{0}\int e^{\frac{\epsilon _{0}}{2}%
p^{2}-\gamma p}dp+\beta e^{\frac{\epsilon _{0}}{2}p^{2}-\gamma p}\overset{%
\infty }{\underset{n=1}{\sum }}\frac{A^{n}}{p^{n}}.
\end{equation*}%
A corresponding form of the Benney hydrodynamic chain is given by%
\begin{eqnarray*}
u_{t} &=&\partial _{x}\left( (\beta A^{1}+\alpha )e^{-\epsilon _{0}u}+\frac{%
\gamma }{\epsilon _{0}}u\right) , \\
&& \\
A_{t}^{k} &=&A_{x}^{k+1}+(-\epsilon _{0}A^{k+1}+\gamma
A^{k}+(k-1)A^{k-1})u_{x}\text{, \ \ }k=1,2,...
\end{eqnarray*}

\textbf{Remark}: Hydrodynamic reductions (\ref{ma}) depend on $N+3$
arbitrary parameters $\alpha ,\beta ,\epsilon _{0}$ and $\epsilon _{k}$,
because the constant $\gamma $ can be removed by the Galilean transformation.

\textbf{2}. $K+2$ component hydrodynamic reduction (cf. (\ref{zak}) and (\ref%
{tri}))%
\begin{equation}
u_{t}=v_{x}\text{, \ \ \ \ }v_{t}=\partial _{x}\left( w-\frac{u^{2}}{2}%
\right) ,\text{\ \ \ \ \ }a_{t}^{k}=\left( \frac{(a^{k})^{2}}{2}+u\right) 
\text{, \ \ }k=1,2,...,K,  \label{na}
\end{equation}%
where $w\equiv H_{2}(\mathbf{a},u,v)$, is a semi-Hamiltonian hydrodynamic
type system if at least one extra conservation law (see (\ref{cona}), $n=2$)%
\begin{equation*}
w_{t}=\partial _{x}\left( s-uv\right) ,
\end{equation*}%
where $s\equiv H_{3}(\mathbf{a},u,v)$, exists. In such a case, the
compatibility conditions $\partial _{i}(\partial _{k}s)=\partial
_{k}(\partial _{i}s)$, $\partial _{k}(\partial _{u}s)=\partial _{u}(\partial
_{k}s)$, $\partial _{k}(\partial _{v}s)=\partial _{v}(\partial _{k}s)$, $%
\partial _{v}(\partial _{u}s)=\partial _{u}(\partial _{v}s)$ lead to the
Gibbons--Tsarev system (cf. (\ref{gt}))%
\begin{eqnarray*}
(a^{i}-a^{k})\partial _{ik}w &=&\partial _{k}w\cdot \partial _{iv}w-\partial
_{i}w\cdot \partial _{kv}w\text{, \ }i\neq k, \\
&& \\
\partial _{k}w\cdot \partial _{vv}w &=&(\partial _{v}w-a^{k})\partial
_{kv}w+\partial _{ku}w, \\
&& \\
\partial _{uu}w+\partial _{v}w\cdot \partial _{uv}w &=&(\partial
_{u}w-u)\partial _{vv}w+\partial _{v}\delta w, \\
&& \\
a^{k}\partial _{ku}w+\partial _{k}w\cdot \partial _{uv}w &=&(\partial
_{u}w-u)\partial _{kv}w+\partial _{k}\delta w,
\end{eqnarray*}%
where $\partial _{u}s=v+\delta w+(\partial _{u}w-u)\partial _{v}w,\partial
_{v}s=u+\partial _{u}w+(\partial _{v}w)^{2}$ and $\partial _{i}s=(\partial
_{v}w+a^{i})\partial _{i}w$. A consistency of (\ref{on}) and (\ref{na})
leads to the L\"{o}wner equations (cf. (\ref{lev}))%
\begin{eqnarray*}
\partial _{i}p &=&\frac{\partial _{i}w}{a^{i}-p}\left( p^{2}-p\partial
_{v}w+u-w_{u}+\sum \frac{\partial _{m}w}{a^{m}-p}\right) ^{-1}, \\
&& \\
\partial _{u}p &=&(\partial _{v}w-p)\left( p^{2}-p\partial
_{v}w+u-w_{u}+\sum \frac{\partial _{m}w}{a^{m}-p}\right) ^{-1}, \\
&& \\
\partial _{u}p &=&-\left( p^{2}-p\partial _{v}w+u-w_{u}+\sum \frac{\partial
_{m}w}{a^{m}-p}\right) ^{-1}.
\end{eqnarray*}%
Suppose that $w$ depends on $u,v$ and a sole function $\Delta =\Sigma
f_{k}(a^{k})$ only. This ansatz reduces the above Gibbons--Tsarev system to%
\begin{eqnarray*}
\partial _{u\Delta }w &=&\partial _{\Delta }w\cdot \partial _{vv}w,\text{ \
\ \ \ \ }\partial _{uu}w+\partial _{v}w\cdot \partial _{uv}w=(\partial
_{u}w-u)\partial _{vv}w, \\
&& \\
\partial _{\Delta \Delta }w &=&0,\text{\ \ \ \ \ }\partial _{v\Delta }w=0,%
\text{ \ \ \ \ \ }\partial _{uv}w+a^{k}\frac{\partial _{u\Delta }w}{\partial
_{\Delta }w}=\frac{f_{k}^{\prime \prime }(a^{k})}{f_{k}^{\prime }(a^{k})},
\end{eqnarray*}%
whose solution is given by%
\begin{eqnarray*}
w &=&(\beta \Delta +\alpha )e^{-\epsilon _{0}u}-\frac{\epsilon _{0}}{2}%
v^{2}+(\gamma u+\delta )v+\frac{\epsilon _{0}-\gamma ^{2}}{2\epsilon _{0}}%
u^{2}+\left( -\frac{1+\gamma \delta }{\epsilon _{0}}+\frac{\gamma ^{2}}{%
\epsilon _{0}^{2}}\right) u, \\
&& \\
\ln f_{k}^{\prime } &=&\ln \epsilon _{k}+\gamma a^{k}-\frac{\epsilon _{0}}{2}%
(a^{k})^{2},
\end{eqnarray*}%
where $\alpha ,\beta ,\gamma ,\epsilon _{0}$ and $\epsilon _{k}$ are
constants.

Then these L\"{o}wner equations can be integrated once, and equation of the
Riemann surface can be found in quadratures%
\begin{eqnarray*}
d\lambda  &=&d\left[ \left( v+\frac{p-\gamma u-\delta }{\epsilon _{0}}+\frac{%
\gamma }{\epsilon _{0}^{2}}\right) e^{\frac{\epsilon _{0}}{2}p^{2}-\gamma
p+\epsilon _{0}u}\right] +\epsilon _{0}(\beta \Delta +\alpha )e^{\frac{%
\epsilon _{0}}{2}p^{2}-\gamma p}dp \\
&& \\
&&-\beta e^{\frac{\epsilon _{0}}{2}p^{2}-\gamma p}\sum f_{m}^{\prime
}(a^{m})d\ln (p-a^{m}).
\end{eqnarray*}%
Introducing moments $A^{k}$ by their derivatives%
\begin{equation*}
dA^{k}=\sum (a^{m})^{k-2}f_{m}^{\prime }(a^{m})da^{m}\text{, \ }k=2,3,...,
\end{equation*}%
the above equation reduces to%
\begin{equation*}
\lambda =\left( v+\frac{p-\gamma u-\delta }{\epsilon _{0}}+\frac{\gamma }{%
\epsilon _{0}^{2}}\right) e^{\frac{\epsilon _{0}}{2}p^{2}-\gamma p+\epsilon
_{0}u}+\alpha \epsilon _{0}\int e^{\frac{\epsilon _{0}}{2}p^{2}-\gamma
p}dp+\beta e^{\frac{\epsilon _{0}}{2}p^{2}-\gamma p}\overset{\infty }{%
\underset{n=2}{\sum }}\frac{A^{n}}{p^{n-1}}.
\end{equation*}%
A corresponding form of the Benney hydrodynamic chain is given by%
\begin{eqnarray*}
u_{t} &=&v_{x},\text{ \ \ \ \ \ }v_{t}=\partial _{x}(w-u^{2}/2), \\
&& \\
A_{t}^{k} &=&A_{x}^{k+1}+(\epsilon _{0}A^{k+1}+\gamma
A^{k}+(k-2)A^{k-1})u_{x}\text{, \ \ }k=2,3,...,
\end{eqnarray*}%
where%
\begin{equation*}
w=(\beta A^{2}+\alpha )e^{-\epsilon _{0}u}-\frac{\epsilon _{0}}{2}%
v^{2}+(\gamma u+\delta )v+\frac{\epsilon _{0}-\gamma ^{2}}{2\epsilon _{0}}%
u^{2}+\left( -\frac{1+\gamma \delta }{\epsilon _{0}}+\frac{\gamma ^{2}}{%
\epsilon _{0}^{2}}\right) u.
\end{equation*}

\textbf{Remark}: Hydrodynamic reductions (\ref{na}) depend on $N+4$
arbitrary parameters $\alpha ,\beta ,\gamma ,\epsilon _{0}$ and $\epsilon
_{k}$, because the constant $\delta $ can be removed by a combination of a
shift and the Galilean transformation.

\section{Conclusion and Outlook}

In the same way, all other hydrodynamic reductions (\ref{j}) can be
completely described. Possibly, most general ansatz is given by (\ref{tri}),
where the function $H_{K}(\mathbf{a},\mathbf{H})$ depend on a \textit{finite}
number of moments $\tilde{A}^{l}$ (see (\ref{reg})), i.e. $l=0,1,...,L$,
where $L$ is some natural number. For instance, let us suppose that $H_{0}$
depends on $\Delta _{1}=\Sigma f_{k(1)}(a^{k})$ and $\Delta _{2}=\Sigma
f_{k(2)}(a^{k})$ only, and $f_{i}(a^{k})$ are unknown functions yet. In the
simplest case, $f_{k(1)}(a^{k})=\epsilon _{k}a^{k}$ and $f_{k(2)}(a^{k})=%
\epsilon _{k}(a^{k})^{2}$. Corresponding hydrodynamic reduction is
determined by $H_{0}=[\Delta _{2}-\Delta _{1}^{2}/(1+\Sigma \epsilon
_{m})]/2 $.

Two important observations are made in this paper. The L\"{o}wner equations
can be derived \textit{faster} via generating functions of commuting flows
(see, for instance, Section 2) than directly from a consistency of
generating functions of conservation laws and corresponding ``symmetric''
hydrodynamic reductions. The Gibbons--Tsarev system can be derived \textit{%
faster} from an existence of an extra conservation law than from a
consistency of the L\"{o}wner equations. It looks like that this is a common
feature of any integrable hydrodynamic chains. However, this, possibly,
universal behavior needs a priori known formulation of hydrodynamic chains
in their conservative form as well as generating functions of commuting
flows, that is usually can be found just after a special separate
investigation.

\section*{Acknowledgement}

I thank Eugeni Ferapontov, John Gibbons, Andrea Raimondo, Vladimir Sokolov
and Sergey Tsarev for their stimulating and clarifying discussions.

I am grateful to the Institute of Mathematics in Taipei (Taiwan) where some
part of this work has been done, and especially to Jen-Hsu Chang for
hospitality in National Defense University. This research was particularly
supported by the Russian--Taiwanese grant 95WFE0300007 (RFBR grant
06-01-89507-HHC).

\end{document}